# Citation Swing Factor: An Indicator to Measure the Diffusion of Cited Items

Bidyarthi Dutta

*Department of Library and Information Science, Vidyasagar University, Midnapore, West Bengal, INDIA.*

**ABSTRACT**
The *h*-index, introduced by Hirsch, is based on the mutual variation between the number of cited and source items. The temporally continuous nature of the citation accretion process causes a shift of cited items from the h-core zone to the adjacent citation-asymmetric zones, viz. *h*-excess zone, or *h*-tail zone. The name given to this shifting phenomenon is the Diffusion of Cited Items (DCI). In this paper, two new variables are introduced, i.e., the Fold of Excess citation over Total citations (FET), denoted by $\varepsilon^2$ and the Fold of h-core citation over Excess citations (FHE), denoted by $\theta^2$. On the basis of $\theta$ and $\varepsilon$, another indicator is introduced, i.e., the Citation Swing Factor (CSF), defined as $\frac{d\theta}{d\varepsilon}$, which indicates the differential coefficient of $\theta$ with respect to $\varepsilon$. The time dependence of FET and FHE is also discussed. The possible solutions of $\frac{d\theta}{d\varepsilon}$ are derived here. The functionality of CSF ($\frac{d\theta}{d\varepsilon}$) to measure the diffusion process quantitatively will be tested later on for journals, authors and institutions.

## INTRODUCTION

The *h*–index, introduced by Hirsch[1] and used by major citation databases, is an author or journal-level metrics. Followed by *h*-index, other associated metrics were subsequently introduced, e.g., *e* index, *R* index,[2,3] *g*-index,[4,5] *et al.* Despite of its effectiveness and simplicity, there are some downsides, such as the loss of citation information and low resolution, resulting from its low potential. There are many papers describing advantages, disadvantages and possible areas of applications of *h*-index.[6-16] Wu[17] and Kosmulski[18] introduced two *h*-type indices, viz. *W*–index and Kosmulski index, which were generalised by Egghe[19] in Lotkaian framework. Fassin[20] introduced h(3) index, another *h*-type index. Waltman and Van[21] discussed various inconsistencies of *h*-index. Also, the *h*-index has a relatively narrow range. For example, in any field, scientists having an *h*-index larger than 100 (which means at least 10,000 citations) are rare. Therefore, due to low resolution, it is quite common for a group of scientists to have an identical *h*-index. This lacuna was resolved by the e-index, which is a real number to complement the *h*-index for the ignored excess citations. Mathematically speaking, the square-root of the number of *h*-core citations, *h*-excess citations and total citations are defined as *h*-index, *e*-index and *R*-index, respectively.

## OBJECTIVES

The objectives of this paper are to introduce a new indicator along with the theoretical foundation on the basis of total number of citations ($R^2$), *h*-core citations ($h^2$) and excess citation ($e^2$). The proposed function of this indicator is to measure the relative change of number of cited items in *h*-tail zone, *h*-core zone and *h*-excess zone. The name given to it is Citation Swing Factor (CSF). The two variables, viz. Fraction of *H*-core citation to Excess citation or FHE ($\theta^2$) and Fraction of Excess citation to Total citation or FET ($\varepsilon^2$), are introduced here to assess the centrifuging and scattering of citations over three *h*–zones respectively. The continuous temporality of the citation accretion process causes a shift of cited items from the *h*-core zone to the adjacent citation–asymmetric zones, viz. *h*-excess zone, which is described here as the diffusion process of cited items. The change of $\theta$ and $\varepsilon$ will be $d\theta$ and $d\varepsilon$ respectively and the relative change, i.e. $\frac{d\theta}{d\varepsilon}$ measures the diffusion, which defines the indicator CSF. The temporal variation of $\theta$ and $\varepsilon$ are also presented on the basis of power law. This indicator will be tested for wide range of journals, authors and institutions to array the spectrum of numerical values in subsequent research.

### Diffusion of Cited Items

The *h*-core citations are confined within a citation–symmetric zone known as *h*-core zone (Figure 1). The continuous temporal variation between cited and source items when meets at a common point in due course, the *h*-index is defined. An equal length (*h*) on both cited and source axes subtends a square of area $h^2$ under the cited vs. source curve, known as the *h*-core zone. The source or citing items and cited items residing within this zone are *h*-core citations and *h*-core papers, respectively. As the citation accretion process is a continuous time-dependent phenomenon, the temporal change of cited vs. source mutual variational pattern causes a shift of cited items from the *h*-core zone to the adjacent citation-asymmetric zones, viz. *h*-excess zone. The name given to this shifting phenomenon of cited items is the diffusion of cited items. In this paper, two new variables are introduced, i.e., the Fold of Excess citation over Total citations (FET), denoted by $\varepsilon^2$ and the Fold of *h*-core citation over Excess citations (FHE), denoted by $\theta^2$, i.e., $\varepsilon^2 = \frac{e^2}{R^2}$ and $\theta^2 = \frac{h^2}{e^2}$. Being ratios of real numbers, $\theta$ and $\varepsilon$ are continuous variables, which may also be endorsed by the resemblance of *h*-index and *R*-index for infinite sequences.[22]

On the basis of these, another indicator is introduced, the name given to which is Citation Swing Factor (CSF), defined as $\frac{d\theta}{d\varepsilon}$, i.e., the differential coefficient of $\theta$ with respect to $\varepsilon$. The *h*-core citation reflects the centrifugal nature of citation, while the excess citation reflects its scattering nature. The FHE indicates the fractional *h*-core citation or the strength of centrifugal citation. Similarly, FET indicates the fractional excess citation or the strength of scattered citations. The usual nature of citation is to hover around the highly cited items, which tells its centrifugal character. This centrifuging process is always associated with scattering of citations, as the peripherals to the high-cited cores receive fewer citations by diffusion from the core. As a result, the high-cited items of the *h*-core, eventually exceeds the symmetric core zone to asymmetric *h*-excess zone. Also, some items from the low-cited or tail zone shift to the core zone. In this way, an incessant shifting process from tail to excess zone, via the core zone continues. The name has been given to such a Tail-Core-Excess continuous shifting process as Diffusion of Cited Items (DCI). The variables $\theta$ and $\varepsilon$ are mutually interdependent variables along with the function of the age of cited item or time individually. The centrifuging of citations may change with scattering of citations and vice versa. The parameter Citation Swing Factor (CSF) indicates FHE's change with respect to FET, i.e., how the citation centrifuging is influenced by citation scattering. As the factor $e^2$ is common in both $\theta^2$ and $\varepsilon^2$, therefore it may be asserted that $\theta = f(\varepsilon)$ and conversely, $\varepsilon = f(\theta)$. The different citation zones are presented in Figure 1.

### Analytical Formalism

The excess citations received by all papers in the *h*-core, denoted by $e^2$, are[2]

$$e^2 = \Sigma(C_j - h) = \Sigma C_j - h^2 \ (1 \leq j \leq h) \ ....(1)$$

Where $c_j$ are the citations received by the jth paper and $e^2$ denotes the excess citations received by the h-core papers. Assuming,

$$d^2 = \Sigma C_j \ .........................................(2)$$

It is obtained[2], $d^2 = e^2 + h^2$; .....(3) Here $e \geq 0$ and e is a real number.

Or $e = \sqrt{(d^2 - h^2)}$ .......(4)

The relationship between *h* and *e*, as expressed in equation (3) and equation (4), instantly depicts a plane spanned by two axes, *h* and *e*, or *h-e* plane. Now, an arbitrary point in the *h-e* plane represents the overall information of citations received by all papers in the *h*-core. It is interesting to point out that the Euclidean distance (*R*) between the origin and the point P(*h,e*) is equal to

$$R = \sqrt{(h^2 + e^2)} = d \qquad (5)$$

The X-axis and Y-axis represent the number of publications and citations, respectively. The area under the rectangular hyperbolic curve (Figure 1) represents the total number of citations received, which is segmentized into three components. The h-core citation is represented by the shaded square ($h^2$) area, while the total excess citation is scattered outside the shaded square area, under the curve (Figure 1) divided into two segments by the $h^2$-zone, viz. upper h-core zone and lower *h*-core zone. The upper and lower h-core zones residing adjacent to Y-axis and X-axis, together represent the number of excess citations over *h*-core citations. The number of citations in the lower *h*-core zone is also known as Tail Citation.[23] Actually, tail citation also belongs to the category of excess citation, but as it consists of a large number of publications received a low number of citations (1, 2, 3 ....), therefore the name 'Tail' resembling trough of the graph. The citations in the upper *h*-core zone are also known as an *h*-core excess citation that distinguishes it from its 'Tail' counterpart. The *h*-core citation indicates the cluster of h–h citations vs. papers, which is the result of the accumulation of 'h' citations over at least 'h' papers. Larger the area of $h^2$ (Figure 1), the value of h-index will be proportionately greater. For any fixed total number of citations, the steady increase in the value of h-core citations ($h^2$) would gradually reduce the excess citation ($e^2$). The h-core citations echo the concentration of citations through clustering over 'h' number of core values. On the contrary, the excess citations portray the scattering of citations outside the h-core or $h^2$ domain. The relative share of h-core citation and excess citation in the corpus of total citation depicts the relative centrifuging and scattering phenomenon of citation over time.

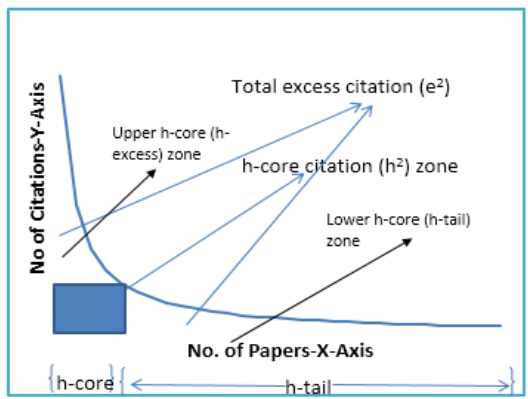

**Figure 1:** Three h-zones (excess, core and tail) in a citation-paper graph.

Now, $\varepsilon^2 = \frac{e^2}{R^2} = \frac{Excess\ Citation}{Total\ Citation} = \frac{Excess\ Citation}{(h-core\ Citation + Excess\ Citation)}$,

so that, $\varepsilon = \frac{e}{R} = \frac{e}{\sqrt{(h^2+e^2)}} = \frac{e}{\sqrt{e^2\left(1+\frac{h^2}{e^2}\right)}} = \left(1+\frac{h^2}{e^2}\right)^{-\frac{1}{2}}$ ..........(6)

Also, $\theta^2 = \frac{h^2}{e^2}$, so, $\theta = \frac{h}{e}$ ..........(7) and substituting in equation (6)

It is obtained, $\varepsilon = (1+\theta^2)^{-\frac{1}{2}}$ ..............(8) (Only positive roots are considered here)

If h-core citation << Excess citation, which may occur for relatively small h-core citation or h-index and high excess citation or e-index, that is to say,

If $h^2 << e^2$ then, $\varepsilon^2 \to 1$, or $\varepsilon \to +1$ (Ignoring '-^' sign)

Now, following theoretically possible cases are analysed here for getting the range of the possible values of θ and ε. An initial assumption is made, that is the minimum possible values of 'P' and 'C' are equal to 1, where P = Number of papers or articles and C = Number of citations.

Case 1: P=1, C=1 → h=1, e=0. It is to be noted that ε can't be equal to 1, as h can't be equal to zero, since the minimum possible value of h-index is 1, for the total number of citation as well as h-core citations equal to 1, resulting e= ε =0, making ε = 0. This is the marginal case, possible only when an author or journal received just one citation for one and only one paper.

Case 2: P=1, C>1, i.e., only one paper receives more than one citation → h=1, e>0, indicating $\varepsilon^2 = \frac{k}{1+k}$, where k represents excess citation $e^2$. If $k >> 1$, then $\varepsilon^2 \approx 1$ and $\varepsilon \to 1$, which can happen if C >>1 or C → ∞, i.e., when the only one paper receives a very large citation.

Case 3: P>1, C=1, i.e., more than one papers receives one citation indicating h=1, e=0, making and consequently ε=0

Case 4: P>1, C>1, i.e., more than one papers receive more than one citation. The following three cases may occur:

Case 4.1: P→∞, C→0, i.e., a large number of papers receive small number of citations indicating $h \to 1$ and $e \to 0$ and making ε→0. If h→l and e →k, where l and k are very low numbers and l ≈ k, then $\varepsilon^2 = \frac{k^2}{l^2+k^2} = \frac{1}{1+\left(\frac{l}{k}\right)^2} < 1$ and ε <1

Case 4.2: P→0, C→∞, i.e., a small number of papers receive a large number of citations indicating h→1, e→∞ (h<<e), making $\varepsilon^2 = \frac{e^2}{(h^2+e^2)} \approx 1$ and also ε≈1.

Case 4.3: P→∞, C→∞, i.e., a large number of papers receive large number of citations. The following three cases may occur:

Case 4.3.1:

h→l and e→∞, i.e., low h-index and high e - index, h<<e where l indicates a very low number. It indicates ε≈1. The area of the shaded square (Figure 1) is small here.

Case 4.3.2:

h→∞ and e→l, where l is a very low number i.e. h>>e, which indicates ε≈0. The area of the shaded square (Figure 1) is large here.

Case 4.3.3: $h \approx e$, i.e., the values of h-core citation and excess citation are in the same order. It indicates, ε≈0.5. The area of shaded square (Figure 1) here is comparable with the area of the upper h-core zone.

An analysis of all cases and sub-cases instantly tells the maximum and minimum values or the range of $\varepsilon^2$ as 1 and 0 respectively, i.e., 0≤$\varepsilon^2$≤1 indicating 0≤ε≤1. Hence $0 \frac{1}{(1+\theta^2)} \leq 1$ (From equation (8)).

For $\varepsilon^2$=0, the value of $\theta^2$ should tend to very large value, or θ→∞, i.e. $\frac{h^2}{e^2}$ should be very large, or the value of $h^2$ (h-core citation) would be much greater than excess citation ($e^2$), i.e., $h^2>>e^2$, or h>>e.

For $\varepsilon^2$=1, $\frac{1}{1+\theta^2} = 1$, indicating $\theta^2$=0. Now, $\theta^2$=0, implies $h^2$=0, or h-core citation is zero, which is not possible as the minimum possible value of $h^2$ is 1. The value of h-index also can't be zero, ($h_{min}$=1). Hence $\theta^2$=0, only when e →∞, or very large value. Actually $\theta^2$ can't be equal to zero, but very close to zero, or $\theta^2$→0. Thus, θ≠0 but θ→0.

In brief, for ε=0, θ→∞; and for ε=1, θ→0, implying the range of θ, as 0 < θ <∞.

Now, $\varepsilon = (1+\theta^2)^{-\frac{1}{2}}$

Now, for 0≤θ≤1,

$(1+\theta^2)^{-\frac{1}{2}} = 1 - \frac{1}{2}\theta^2 + \frac{\left(-\frac{1}{2}\right)\left(-\frac{3}{2}\right)}{2!}(\theta^2)^2 + \frac{\left(-\frac{1}{2}\right)\left(-\frac{3}{2}\right)\left(-\frac{5}{2}\right)}{3!}(\theta^2)^3 + \cdots$

$= 1 - \frac{\theta^2}{2} + \frac{3}{8}\theta^4 - \frac{5}{16}\theta^6 + \ldots \approx 1 - \frac{\theta^2}{2}$ (By applying Maclaurin's expansion theorem and neglecting small quantities of higher order)

Hence, $\varepsilon = 1 - \frac{\theta^2}{2}$, for $\theta < 1$, or $\theta = \sqrt{2(1-\varepsilon)}$ (Taking positive root only)

If the two variables (ε,θ) characterizing the two states, viz. scattering of citations (FET) and centrifuging of citations (FHE), differ from each other only infinitesimally, the change of $\theta$ and $\varepsilon$ will be $d\theta$ and $d\varepsilon$ respectively. Differentiating both sides of equation (8) with respect to ε, it is obtained,

$\frac{d\theta}{d\varepsilon}$ = Citation Swing Factor (CSF) = $-\frac{1}{\theta\varepsilon^3} = -\frac{R^3}{he^2}$ (for $\theta^2 > 1$, or $h^2 > e^2$, i.e., h index > e index)..............(9A)

[Since, $\varepsilon^2(1 + \theta^2) = 1$

$\therefore \theta^2 = \frac{(1-\varepsilon^2)}{\varepsilon^2}$, or $\theta = \sqrt{\frac{(1-\varepsilon^2)}{\varepsilon^2}} = \sqrt{\left(\frac{1}{\varepsilon^2} - 1\right)} = \left(\frac{1}{\varepsilon^2} - 1\right)^{1/2}$ .....................(8A)

Differentiating both sides of equation (8A) with respect to θ, we get

$$\frac{d\theta}{d\varepsilon} = -\frac{1}{\varepsilon^3}\left(\frac{\varepsilon^2}{1-\varepsilon^2}\right)^{1/2} = -\frac{1}{\varepsilon^2\sqrt{(1-\varepsilon^2)}}$$

Now, by basic definition, $\varepsilon = \frac{e}{R}$ and $\theta = \frac{h}{e}$; (By equation (6) and (7)) and $h^2+e^2=R^2$, By equation (5). Thus,

$$\frac{d\theta}{d\varepsilon} = -\frac{1}{\frac{e^2}{R^2}*\sqrt{\frac{(R^2-e^2)}{R^2}}} = -\frac{R^3}{he^2}\,]$$

Also, $\frac{d\theta}{d\varepsilon} = -\frac{1}{\theta} = -\frac{e}{h}$ ... (for $\theta < 1$, $h^2 < e^2$, i.e., h index < e-index)......(9B), by substituting $\varepsilon$ and $\theta$ from equations (6) and (7)

This expressions for CSF (eq. 9(A) and 9(B)) have been derived on the basis of the mutual interdependence between $\theta$ and $\varepsilon$, i.e., $\theta=f(\varepsilon)$ and $\varepsilon=f(\theta)$. The time factor has not been so long included. Now, as both $\theta$ and $\varepsilon$ are also functions of age of cited item or time, it may be expressed:

$\theta=f(\varepsilon,t)$...(10) and $\varepsilon=f(\theta,t)$...(11).

If the two variable-pairs [(ε,t)and (θ,t)] characterizing the two states, viz. temporal scattering of citations and temporal centrifuging of citations, differ from each other only infinitesimally, the change of $\theta$ and $\varepsilon$ will be $d\theta$ and $d\varepsilon$ respectively. Now, $d\theta$ and $d\varepsilon$ are exact differentials (for, e>0 and R>0), because $\theta$ and $\varepsilon$ are well-defined functions, except at e=0 and R=0. Hence, $d\theta = \left(\frac{\delta\theta}{\delta\varepsilon}\right)_t + \left(\frac{\delta\theta}{\delta t}\right)_\varepsilon$.......(10A) and $d\varepsilon = \left(\frac{\delta\varepsilon}{\delta\theta}\right)_t + \left(\frac{\delta\varepsilon}{\delta t}\right)_\theta$........(11A), by applying partial differentiation to both sides of equation (10) and (11) respectively. Now, $\left(\frac{\delta\theta}{\delta\varepsilon}\right)_t \equiv \frac{d\theta}{d\varepsilon}$ and $\left(\frac{\delta\varepsilon}{\delta\theta}\right)_t \equiv \frac{d\varepsilon}{d\theta}$ and assuming the temporal dependence of $\theta$ and $\varepsilon$, in accordance with the power law[2], i.e., $\theta=\theta_m t^{-k}$ ($t\geq 1$, $k>0$).....(12), where $\theta_m$ is the maximum value of $\theta$ at t=1 and $\varepsilon= \varepsilon_m t^{-l}$ ($t\geq 1$, $l>0$)......(13) where $\varepsilon_m$ is the maximum value of ε at t=1, k and l are constants, it is obtained,

$$d\theta = -\frac{R^3}{he^2} - \frac{\theta_m k}{t^{k+1}}......(14) \text{ and}$$

$$d\varepsilon = -\frac{he^2}{R^3} - \frac{\varepsilon_m l}{t^{l+1}}.........(15), \text{ for } h^2>e^2 \text{ and}$$

$$d\theta = -\frac{e}{h} - \frac{\theta_m k}{t^{k+1}}......(14A) \text{ and}$$

$$d\varepsilon = -\frac{h}{e} - \frac{\varepsilon_m l}{t^{l+1}}........(15A), \text{ for } h^2<e^2.$$

As time passes, t will be larger and consequently, the terms containing 't' will tend to zero, while the first terms will outshine.

## CONCLUSION

The concepts of citation centrifuging and citation scattering are presented here in terms of two continuous variables FHE (θ) and FET (ε), while another new indicator CSF is defined as $\frac{d\theta}{d\varepsilon}$, to envisage the process of diffusion of cited items across the tail–core–excess boundaries. The temporal dependence of θ and ε are discussed. The theoretically possible solutions of CSF are analysed here. The practical applicability of CSF will be tested for journals, authors and institutions subsequently.

## ACKNOWLEDGEMENT

This work is executed for the development of the theoretical foundation of the research project entitled Design and development of comprehensive database and scientometric study of Indian research output in physics and space science since independence sponsored by Department of Science and Technology, Govt. of India under NSTMIS scheme, (Vide F. No. DST/NSTMIS/05/252/2017-18 dated 11/01/2018). The scholarly discussions with Dr. A. N. Rai (DST), Dr. P. Arora (DST) and Dr. B. K. Sen added worth to this paper that are hereby gratefully acknowledged.

## CONFLICT OF INTEREST

The author declares no conflict of interest.

## ABBREVIATIONS

**CSF:** Citation Swing Factor; **FHE:** Fraction or Fold of *h*-core citation to Excess Citation; **FET:** Fraction or Fold of Excess Citation to Total Citation; **DCI:** Diffusion of Cited Items.

## REFERENCES

1. Hirsch JE. An Index to Quantify an Individual's Scientific Research Output. Proceedings of the National. 2005;102(46):16569-72. arXiv:physics/0508025
2. Zhang CT. The e-Index, Complementing the h-Index for Excess Citations. PLoS One. 2009;4(5):e5429. doi:10.1371/journal.pone.0005429
3. Jin BH, Liang LM, Rousseau R, Egghe L. The R- and AR-indices: Complementing the h-index. Chinese Science Bulletin. 2007;52(6):855-63.
4. Egghe L. An Improvement of the h-index: The g-index. ISSI Newsletter. 2006;2(1):8-9.
5. Egghe L. Theory and Practise of the g-index. Scientometrics. 2006;69(1):131-52.
6. Egghe L. The Hirsch Index and Related Impact Measures. In Cronin (Ed.), An-

nual Review of Information Science and Technology. Medford, NJ, USA: Information Today, Inc. 2010;65-114.

7. Egghe L. Distributions of the h-index and the g-index. In Torres-Salinas and amp; H. F. Moed (Eds.), Proceedings of ISSI 2007. 11th International Conference of the International Society for Scientometrics and Informetrics. 2007;245-53.

8. Egghe L. The Influence of Transformations on the h-index and the g-index. Journal of the American Society for Information Science and Technology. 2008;59(8):1304-12.

9. Egghe L. Mathematical Theory of the h-index and g-index in case of Fractional Counting of Authorship. Journal of the American Society for Information Science and Technology. 2008;59(10):1608-16.

10. Egghe L. Examples of Simple Transformations of the h-index: Qualitative and Quantitative Conclusions and Consequences for other Indices. Journal of Informetrics. 2008;2(2):136-48.

11. Egghe L. The Influence of Merging on h-type Indices. Journal of Informetrics. 2008;2(3):252-62.

12. Egghe L. Conjugate Partitions in Informetrics: Lorenz-Curves, h-type Indices, Ferrers Graphs and Durfee Squares in a Discrete and Continuous Setting. Journal of Informetrics. 2010;4(3):320-30.

13. Egghe L. Modelling Successive h-indices. Scientometrics. 2008;77(3):377-87.

14. Egghe L. Mathematical Study of h-index Sequences. Information Processing and amp; Management. 2009;45(2):288-97.

15. Egghe L. Comparative Study of h-index Sequences. Scientometrics. 2009;81(2):311-20.

16. Egghe L. Performance and its Relation with Productivity in Lotkaian Systems. Scientometrics. 2009;81(2):567-85.

17. Wu Q. The w-index: A Measure to Assess Scientific Impact by Focusing on Widely Cited Papers. Journal of the American Society for Information Science and Technology. 2010;61(3):609-14.

18. Kosmulski M. A New Hirsch-type Index Saves Time and Works Equally Well as the Original Index. ISSI Newsletter. 2006;2(3):4–6.

19. Egghe L. Characterizations of the Generalized Wu- and Kosmulski-Indices in Lotkaian Systems. Journal of Informetrics. 2011;5(3):439-45.

20. Fassin Y, Rousseau R. The h(3)–Index for Academic Journals. Malaysian Journal of Library and amp; Information Science. 2018;24(2):41-53.

21. Waltman L, Van ENJ. The Inconsistency of the h-index. Journal of the American Society for Information Science and Technology. 2012;63(2):406-15.

22. Egghe L, Rousseau R. Infinite Sequences and their h-type Indices. Journal of Informetrics. 2019;13(1):291-8.

23. Baum JA. The Excess-Tail Ratio: Correcting Journal Impact Factors for Citation Distributions. Management. 2013;16(5):697-706.